# U-LEARNING WITHIN A CONTEXT-AWARE MULTIAGENT ENVIRONMENT

Monica Vladoiu[1] and Zoran Constantinescu[2]

[1]Department of Informatics, PG University of Ploiesti, Ploiesti, Romania
`monica@unde.ro`
[2]Department of Research and Development, ZealSoft Ltd., Bucharest, Romania
`zoran@zealsoft.ro`

## ABSTRACT

*New technological developments have made it possible to interact with computer systems and applications anywhere and anytime. It is vital that these applications are able to adapt to the user, as a person, and to its current situation, whatever that is. Therefore, the premises for evolution towards a learning society and a knowledge economy are present. Hence, there is a stringent demand for new learner-centred frameworks that allow active participation of learners in knowledge creation within communities, organizations, territories and society, at large. This paper presents the multi-agent architecture of our context-aware system and the learning scenarios within ubiquitous learning environments that the system provides support for. This architecture is the outcome of our endeavour to develop ePH, a system for sharing public interest information and knowledge, which is accessible through always-on, context-aware services.*

## 1. INTRODUCTION

Context-awareness refers to the ability of a class of systems to use contextual information to supply better services to the user, in a flexible and manageable way. Context can be seen as "any information that can be used to characterize the situation of an entity", where an entity can lay within a large range of things: a person, a place, an object, etc., that is relevant to the interaction between a user and an application [1]. A context-aware system is able to extract, interpret and use contextual information and to adapt its functionality to the current context of use in order to provide the right services to particular people, place, time, event, etc. [2].

When considering learning in this perspective, we see that it is no longer an isolated endeavour, but more of a social process that relies on collaboration and communities. This new learning paradigm is possible within dynamic contexts, where the learner acquires knowledge and skills in a process that involves mutual trust, shared interests, common goals, commitments, and obligations, exchanging of services, and, moreover, a genuinely proactive, motivated attitude towards learning. Therefore, the learners will be learning together, providing added value, sharing and executing tasks in order to reach a common goal. More often, the communities will establish their own goals in terms of knowledge and skills to be acquired, instead of using a predefined curriculum. The goals themselves will fully correspond to learners' needs, and will be highly dependent on the local culture and its priorities [3].

Technological progress has seeded the stimuli for evolution towards this new learning society and knowledge economy, in which there is a need to rethink education, training and learning to subscribe to a new learning environment paradigm that provides for lifelong and life-wide learning. In this environment, learners are not simply consumers of knowledge; they (socially) create knowledge for themselves, for their organizations and for their communities. A 21st century learner is able to extracts knowledge from everyday activities, at work and in the community, *through mainly informal, non-formal, intentional, incidental or accidental learning activities, and shares his/her learning with peers within relevant communities in multi-dimensional learning spaces* [4]. Therefore, there is a stringent demand for new learner-centered frameworks that allow active participation of learners in knowledge creation, within learning communities, learning organizations, learning territories (municipalities, regions), and society, in general.

Support for this need comes from ubiquitous computing, which is seen as the process of seamless integration of computers into the real world. Thus, Weiser has pointed out that the *most profound technologies are those that disappear. They weave themselves into the fabric of everyday life until they are indistinguishable from it* [5]. The old desktop computers are quickly replaced by computing resources that are embedded in physical objects of everyday life, by which these are not detracted from their original functionality but enhanced with computing capabilities [6]. Besides this kind of phyical disappearance, ubiquitous computing also generates transparency, i.e. the computers can still be there but the user perceives them as windows to virtual worlds in the new emerging Ubiquitous Learning Environments (ULEs). Therefore, ubiquitous computing provides for ubiquitous learning (u-learning). More specifically, *u-learning that employs mobile devices, wireless communications and sensor technologies in learning activities is called "context-aware u-learning"* [7]. The context-aware feature of the u-computing environments allows better understanding of both the user and the specific situation "around" him or her [7]. The term "ubiquitous" does not refer here simply to anytime/anywhere, but more specifically to the ability *to support multiple diverse learning contexts and automatically adapt to them* [3]. Currently, ubiquitous learning takes place in various education environments, in contrast with early applications of ubiquitous learning (tourist and museum guides), which offered the possibility of getting information based on visitor's current position, e.g. facts about a painting or a tourist attraction she was standing in front of [6].

The main characteristics of ubiquitous learning are as follows: *permanency* (the learners will never lose their work unless it is voluntarily deleted; moreover, all the learning processes are recorded on a daily basis, which allows for later reflection on the learning process), *accessibility* (the learning content is accessible anywhere), *immediacy* (the instant access to the content allows learners to store and retrieve it at anytime), *interactivity* (the learners can interact with facilitators or peers, both synchronously or asynchronously), *situated-ness of the instructional activities* (the learning occurs naturally in the daily life in a context-aware way), *adaptability* to the learner's current situation, both the cyber world and the real world, which provides for personalized active learning), and *non-intrusiveness* (the ubiquitous technology should be as invisible as possible, which should result in natural interaction with the user and, consequently, seamless learning) [3, 5, 7, 8, 9, 10, 11, 12, 13].

This paper presents the multi-agent architecture of our context-aware system (called ePH) and the learning scenarios within ubiquitous learning environments that the system provides support for. This architecture is the outcome of our endeavor to develop ePH, which is a system for sharing public interest information and knowledge that is accessible through always-on, context-aware services [15, 16]. ePH is an acronym for "e-Prahova", where Prahova is the name of our county, for which the system has been developed.

The rest of this paper is structured as follows: the next section contains the related work on context-aware multi-agent system with focus on some systems that have been utilized in real world experiments. The third section presents some context-aware ubiquitous learning environments that are similar in their learning approach with ePH. Section IV describes briefly our context-aware system and illustrates its use within learning scenarios. The ePH's learning capabilities are demonstrated in Section 5. The multi-agent architecture of the system and a possible learner scenario are presented, respectively, in the next two sections. The last section briefly concludes the paper, and points out some future work ideas.

## 2. CONTEXT AWARE MULTI-AGENT SYSTEMS

A multi-agent system is composed of multiple interacting intelligent agents. Multi-agent systems can be used to solve problems, which are difficult or impossible for an individual agent. In [17] the following major characteristics of multiagent systems are identified: each agent has just incomplete information and it is restricted in its capabilities, system control is distributed, data is decentralized, and computation is asynchronous. There is evidence in the literature that context-aware systems are a very active area of research, from which context-aware multi-agent systems covers more than one third [18, 19]. Within this section, we focus on some relevant systems that have been used in real world experiments.

X.MAS is a generic multiagent architecture that is focused on information retrieval tasks that are aimed to retrieve, filter and reorganize information according to user's interests [20]. It consists of the following categories of software agents: *Information agents*, tailored to extract and handle information while accessing information sources, *Filter agents*, which transform information according to user preferences, *Task agents*, able to help users to perform tasks typically in cooperation with other agents, *Interface agents*, in charge of interacting with the user, and *Middle agents*, for establishing communication among requesters and providers. The authors also present some examples of actual systems that have been built based on the proposed architecture – e.g. WIKI.MAS that classify Wikipedia contents according to a predefined set of classes belonging to the WordNet domains.

Reference [21] proposes a Context-aware Service Platform (a prototype system) that utilizes Semantic Web technologies to analyze the ambient contexts and contrive service plans. They have integrated ontology with rule-based reasoning to automatically infer high-level contexts and deduce a goal of context-aware services. An AI planner decomposes complex services and establishes the execution plan, while agents perform the tasks to accomplish the service.

LOQOMOTION is a mobile agent-based architecture for distributed processing of continuous location-dependent queries that are issued by mobile users, which allows the retrieval of objects' locations and other interesting data. LOQOMOTION is able to process queries that depend on the location of any moving object, as opposed to other approaches that, for example, only focus on queries about static objects that are in the vicinity of a certain static or moving object, objects moving within a fixed region, etc. [22]. LOQOMOTION is based on a layered hierarchy of mobile agents that move autonomously over the network with the goal of tracking efficiently the relevant moving objects, correlating partial results, and, finally, presenting and continually updating the answer to the user's query [23].

Cyberguide is one of the first systems to provide for mobile context-aware tour guide. Knowledge of the user's current location, as well as a history of past locations, are used together to provide the kind of services that are expected from a real tour guide [24]. Reference [25] presents a model and a multi-agent architecture that has been designed to solve problems in web domains through the integration of information gathering and planning techniques. A particular implementation (MAPWEB-ETOURISM) has been applied to a specific domain, e-tourism, to solve travel problems.

The AmbieSense system has a multi-agent architecture for context-aware information services for mobile users that consists of wireless context tags, content provision platforms and mobile users [26]. The goal of the AmbieSense project is to develop a platform that satisfies the information needs of travellers and mobile users by providing ambient and personalized information in a context-aware manner.

## 3. UBIQUITOUS LEARNING IN UBIQUITOUS LEARNING ENVIRONMENTS

During the current decade, there has been a real boom of ubiquitous learning projects around the world. Further on, we summarize briefly some of the most similar with our system, from the viewpoint of the learning approach.

MOBIlearn is *a worldwide European-led research and development project that has explored context-sensitive approaches to informal, problem-based and workplace learning by using key advances in mobile technologies*. The user is provided with personalised and location-based information, by combining both her user profile and her current location [11, 27].

The Street Poet project offers digitally augmented and location-aware content in literature education. The Client program that runs on a handheld device obtains the user's GPS position, and use it to playback the relevant education content, allowing users to interact with local artifacts or areas of interest [11, 28]. Similarly, a context aware ubiquitous learning platform that applies GPS and image recognition technologies to support learners while learning about natural herbs, in the real world, it is illustrated in [29].

Another typical application of context aware ubiquitous learning allows language learning, students being provided with the right vocabulary in the current contextual situation. An interesting attempt that integrates support for personalised and contextualised learning has been presented in [12], where a system for learning English in real-world situations, called CLUE, is described. CLUE adapts the learning content to both learner's interest and location, by oferring him the suitable English expressions that denominate the objects in his surroundings [11, 12]. A system for learning the appropriate polite Japanese expression according to a particular context is presented yin [11, 14]. A similar system, called TANGO, detects the objects around the learner based on RFID tags, and assigns questions to the learner related to the detected objects to improve her vocabulary knowledge. Moreover, the learners may share their knowledge in the current context [30]. In [31], the authors propose a u-learning framework based on three context interactions, namely recording context, triggering application action on detected context, and augmenting digital data with context information. Several scenarios of learning Kanji language illustrate the benefits of using u-learning in real everyday life.

A context aware ubiquitous learning environment for peer-to-peer collaborative learning is presented in [32]. It includes three systems, i.e. peer-to-peer content access and adaptation system, personalized annotation management system, and multimedia real-time group discussion system. A collaborative learning scenario having the subject *One week vacation in New York City* is also discussed in the paper. PhotoStudy is another project that allows collaborative use of images and annotation of learning content with images or audios recorded on mobile devices [11, 33].

Another appealing approach is taken in the Musex system, which integrates both formal and informal learning. Through Musex, children can collaboratively challenge some questions which are related to the exhibitions in a Japanese science museum [34]. Thus, the system enhances face-to-face discussion by using PDAs to inform two paired learners about the correctness of their answers to a certain question; hence, it integrates formal learning content with an informal discussion [11].

A similar system is MILE that supports both classical and virtual learning environments, allowing students to collaborate, share notes, participate in whiteboard sessions, etc. [35]. A related approach has been taken in a project that tests the RAFT concept (Remote Accesible Field Trips) that allows the students to participate lively in field trips from remote locations. In fact, only a small number of students are at the field location, while most of their peers are in the classroom. The field and the classroom are connected by means of a number of technological devices that relay data via a high-speed internet connection. The two studies reported in the paper indicate that RAFT provides an engaging and motivating learning experience for students [11, 36].

A challenging approach is taken in [37], where it is introduced a ubiquitous learning environment that integrates the use of traditional books, mobile phones, and Web-based discussion forums to boost the acquisition of knowledge. Thus, the students receive contextual messages from an online learning community according with their learning status. The timely recommendatory messages enable smooth collaboration among the members of the learning community.

## 4. THE EPH SYSTEM

ePH stores regional information and knowledge in ePH-DLib, a user-centered core digital library. Its content is available through always-on context-aware services. Users can get it or augment it no matter where they are: at home or office by using a computer, on road with a specific GPS-based device in the car (called gipix, developed in-house), or off-line/off-road via mobile phone. The users can act as members of a social network or individually. The digital library contains public interest information (drugstores, hospitals, general stores, gas stations, weather, entertainment spots, restaurants, travel and accommodation, routes etc.), historical, touristic, and cultural information and knowledge, users' personal "war stories" (tracks, touristic tours, impressions, photos, short videos and so on), and their additions, comments or updates to the content. This content is available to the ePH's users in a context-based way. For example, for a traveller being in a given area, the system may suggest several locations to go to (and actions to execute to reach them): a restaurant to dine at, a museum or memorial house to visit etc. Moreover, if a user is interested in something in particular, like haunted castles, and s/he is located near such a place, and s/he can reach it within a reasonable time frame (having time to get back before dark), the system could show the tasks to be executed to guide her to reach that place [38]. The system has the ability to help also in a scenario that takes place in a remote region, in which the fuel is going down rapidly - ePH shows on the car's GPS-based device where the nearest gas station is.

Moreover, one important capability of the system is its support for learning in ubiquitous learning environment. The basic learning scenarios involve the possibility to access whatever is relevant to education within a given almost circular area (having a certain radius) or along a particular segment of a track (having a particular length), both in the real and virtual world. The system can support users to fulfill their specific learning goals in a context-aware fashion, by recommending what is worth to be visited (from a learning viewpoint) within the specified area and by displaying the tasks to be executed to guide the user to get to that place.

ePH's architecture incorporates the Communications Server, the Location Server, the CBR Reasoner, the Knowledge Base, the Context Middleware, and the Multi-Agent Action

Subsystems [16]. The *Communications Server* provides the support for the always-on kind of service, regardless of the place where the user is when s/he needs that service. The *Location Server* makes available the right service according to the location. The main tasks of the *CBR Reasoner* are as follows: identification of the current problem situation, retrieval of a past case that is similar to the one in progress, suggestion of a solution to this problem that uses that similar case, evaluation of this solution, and update of the system by learning from the current experience. The *Knowledge Base* incorporates general domain-dependent knowledge and specific knowledge (that is embodied by cases), which are used together to find the solution to a specific user problem (that defines the ePH's architecture as being *knowledge-intensive*).

The *Context Middleware* provides a generic context management infrastructure that gathers and maintains contextual information, freeing the agents and the applications of this chore. Throughout this work we have considered the context definition from [26]: *context is a set of suitable environmental states and settings that concern a user, which are relevant for a situation-sensitive application during the process of adapting the services and the information that is offered to the user*. Once the current context changes, the new context activates the *Multi-Agent Sub-system*, which contains various agents that deal with the context, the CBR process, the task facilitation and decomposition, and the application-specific undertakings [38]. These agents will be described in more detail in Section V. As the digital library, ePH-DLib, can be used both on- and off-line with ePH, it is not seen as strongly connected within this architecture.

The current stage of the project is as follows: the geospatial engine unde.ro provides the basic ePH functionality [15], the car device is in current use, and the vital cores of both the CS and the LS are functional as well. Some experimental results are also available [38, 39]. Currently, we are working on the CBR reasoner, the knowledge base and the context middleware.

## 5. LEARNING WITH EPH

Nowadays, learning is an active process that is blended into our daily life. To enable this kind of learning, the revision of the old educational frameworks is required (moving from *contents* to *context*) to provide for u-learning that takes place *anywhere and anytime and across multiple learning contexts* [11]. Some of the learning opportunities occur in a formal context, while others happen in an informal setting. Therefore, context aware u-learning infrastructures need to integrate both formal and informal learning support [11].

Placing the learner at the centre, in this context, does not mean that s/he is the center of attention of well-meaning teachers and instructors, but the centre of production of knowledge that occurs in various contexts within multi-dimensional learning spaces. Thus, the *learner is now at the center of a multidimensional learning space through an active participation in learning communities (personal and professional), learning organizations, learning territories (municipalities, regions) and society* [4]. The major reason for such a revision remains the massive shift from an industrial society to a knowledge economy and learning society.

As shown briefly in the section that describes our system, ePH has a significant potential to support learning in multi-dimensional learning spaces and in ubiquitous learning environments. There are two relevant learning scenarios, each of them involving the opportunity to access whatever is relevant to one person's educational interest within a given (real or virtual) area. First one is confined inside an almost circular area, while the second takes place along a particular segment of a track. The system help users to fulfil their specific learning goals in a context-aware fashion, by making recommendations on what is worth to be seen within the specified area, from a learning point of view, and by showing the tasks to be executed to guide the user to reach those specific Points of Interest (POIs). Thus, the learner is supported through her knowledge journey smoothly, while she is moving in the environment, with ePH holding her hand all the way to the envisaged destination, so that she remains unconcernedly engaged in the learning process.

The context, as a concept, it is intimately related with reasoning and cognition in humans. For context-aware systems, the context works as a tool to select the correct action, in an action-oriented view of intelligence [40]. There are many works that report on development of context models to be used in problem solving. Thus, the context has been used to improve the quality and efficiency of case-based reasoning within diagnostic domain. The authors had focused on the entwined relationship between the agent doing the reasoning and the characteristics of the problem to be solved [41]. In [42] the focus is on how the agents understand situations relying on the information they can perceive. The concept of context in u-learning has been seen as *any information that describes partially the subjective, physical and social situation of the learner depending on the application needs* [31].

In this work we use a multidimensional context model that subscribes to a meronomy that articulates various works from the literature [15, 16, 21, 22, 23, 24, 47, 48, 49] and it is enriched to fulfil ePH's specific functionality. Thus, the context is multidimensional within the ubiquitous environment, having the following facets:

- *personal context*: *user's interests and intentions* (both general and current), *state of mind, feeling and emotions* - e.g focused or distracted, bored, tired, etc.), *knowledgeability* (education, profession, expertise etc.), *limitations* (health issues, disabilities etc., preferences - e. g. the preferred stimuli: visual, auditory, kinaesthetic), *social customs and cultural habits* (being punctual, openess, friendliness, getting up late in the morning etc.), *motivation* (high, medium or low), *social abilities* (leadership, teamwork, communication, empathy etc.), *cognitive abilities* (alternating, divided, focused, or selective attention, conceptual reasoning, visual tracking etc), *learning style* (activist, reflective, theorist, pragmatic), *objectives and goals* (acquiring new knowledge, practical or transferrable skills, or change attitudes, viewpoints, feelings etc.), and so on;
- *task context:* operations, goals, operating mode – static or dynamic, etc;
- *device context:* mobile phone, gipix, PDA, laptop, desktop etc.;
- *social context:* friends, family, colleagues, acquaintances etc.;
- *spatio-temporal context:* date, time, user's location, orientation and movement, space – e.g. public, private, limitations – e. g. time interval, location area, etc;
- *environmental context* (things, persons, services, weather, indoor/outdoor, illumination, noise, crowded etc. from user's surroundings);
- *user interface*: textual, graphical, 3D, web-based, resolution, dimensions, versatility, etc.;
- *infrastructure*: network related (availability, bandwidth, stability, price, and so on), or other resources related (coverage, battery, charger etc.);
- *strategic context:* something important for a planned effect;
- *historical context:* for keeping trace of the past experience.

These all relate to where the user is, when s/he is using the service, what s/he is using the service for, who s/he is with, what s/he likes etc. However, considerations such as how young the user is, or whether it is raining can be equally important.

The problem can be defined as "take a point from the multi-dimensional context space and map it to a learning point cloud within the space of the intertwined subject circles" (Figure 1a).

Various things of interest from a historical point of view are situated on the historical axis (H): monuments, theme museums (World War Two, French Revolution etc.), buildings (castles, house of parliament etc.), locations (e. g. Austerlitz field) and so on. Geographically relevant information that can be found on the geographical axis (G) is as follows: locations, caves, waterfalls, volcanoes, rivers, deltas, seas, oceans, mountains, natural reservations etc. On the Natural Sciences (NS) axis one can find elements of the flora and fauna (e.g. a botanic garden or a natural reservation), geological formations, astronomy phenomena and so forth. The culture axis (C) includes museums, memorial houses, monuments, cultural events, traditional fairs, workshops and expositions, significant constructions and so on. As many learning points can be associated with more than one axis, the users' interest is generally focused on a *learning point cloud* (Figure 1b).

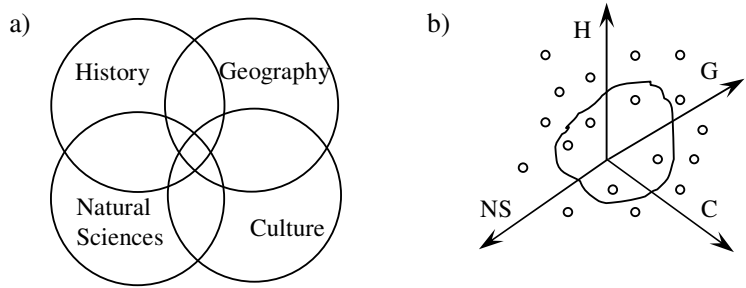

Figure 1. a) The intertwined subject circles b) The learning point cloud

Our solution to this problem is to use case-based reasoning and multi-agent actions subsystems within a knowledge-intensive architecture, as it can be seen in the next section.

## 6. EPH MULTIAGENT ARCHITECTURE

In this section, the multiagent architecture of ePH is presented in more detail (Figure 2). We have called this architecture "FACE" to emphasize our aspirations to provide a knowledge-intensive reasoning process that is inspired by the way in which humans solve problems. A more detailed description of FACE can be found in [38].

Users may benefit from using ePH either on request or event-triggered by a change in the current context. The users can act individually or within a ePH-based social network, both situations being covered by *context widgets* that are able to acquire particular context information and to make it available to the context-aware applications [26]. Once a new context is identified, the CBR Reasoner tries to retrieve a known case and to classify the new situation relying on this case, if possible. After the successful classification of the current situation is performed, the new case will be stored in the case-base as a tuple that includes the contextual information that describes the situation, the problem that corresponds to this situation, and the constructed solution (the retrieved information and knowledge and the tasks to be executed when appropriate). The CBR Reasoner of ePH integrates the classical CBR cycle (Retrieve, Reuse, Revise, Retain) [45] with other reasoning paradigms (rule-based systems, model-based reasoning, deep models – like causal reasoning etc.), as well as with other methods of generating knowledge (data-, text- or knowledge-mining) [38]. The knowledge base incorporates general domain knowledge and case-specific knowledge.

Beside general domain and case knowledge, the knowledge base contains the context model (briefly presented in the previous section), and the cases (the *initial cases* - pre-classified situations that have been acquired prior to first execution, the *point cases* - which are generated to incorporate a new occurent situation, and the *prototypical cases* - generalized cases, that

International Journal of Computer Networks & Communications (IJCNC) Vol.3, No.1, January 2011

contain aggregation of knowledge from previous point cases). Moreover, we need to keep within the knowledge base the user stereotypes and their specific triggers, to support the construction of *user models* [50].

Context templates are used to generate context patterns, and valid context instances are context instantiations that comply with the patterns laid out by the templates. When a new context instance occurs, the Context Agent gets it via the Context Middleware, translates it based on the specific ontology and sends it to the CBR Agent, which tries to identify the current situation (by using the ontology for reasoning about concepts). If this new situation cannot be classified above a given similarity threshold, then a new case is created by the New case Creator Agent. Further, the Task Decomposer is notified about this particular situation. The notification contains the situation, the corresponding contextual information, the task description and the associated goal. The Task Decomposer Agent separates the task into the required sub-tasks and recruits the correct application agents for solving those tasks. There is a large variety of such agents: device agent, GIS agent, reasoning agents, knowledge mining agent, information retrieval agent etc. We describe some of these agents below.

The Device Agent has three functionalities. The first one is to configure the device's applications and to present the prepared information. Secondly, it tries to obtain the user identification and all the possible data about the user's current context. The third functionality is to communicate with the system by using any appropriate means of communication. The device agent has different implementations for different types of devices (e.g. cars, mobile phones, PDAs, notebooks or desktop PCs). It contains more information about the device's specific hardware capabilities, and it can obtain additional information about available software tools.

The GIS Agent encapsulates spatial analysis and query. It can respond to outside requests, and carries out specific spatial analysis and queries that have different data requests. Spatial data is accessed through one ore more GIS servers, to which it may interface. This enables the distribution and manipulation of maps, models, tools and specific points of interest within the ePH framework, in a way that fits well to users' needs. The GIS Server allows the developers to author cost-effective maps, points of interest and geo-processing tasks.

The Information Retrieval agent is responsible for extracting data from different information repositories (Internet web sites, digital libraries, other trusted sources etc.). Each information agent is associated with one information source, the extraction of data being done using specialized wrappers, which can also transform the relevant information into a structured form, if necessary.

## 7. EPH USER SCENARIO

Further on we present the way in which a user who is interested in learning experiences can benefit from interaction with ePH. The idea behind this kind of support has been to help a person who is at a given time in a certain location to experience as most as possible as a learner, in a personalized and effective way, both in the real and virtual world.  Let us consider a scenario, in which Tudor, an ePH user, is interested in learning about the specificity of the Byzantine architecture of the monasteries in our county. Tudor may be planning his learning trip prior to the journey itself or he might adjust his excursion dynamically, as he gets close to some POIs that are relevant to him. The static scenario is illustrated in Figure 3, where one can also see the related prototypical case.

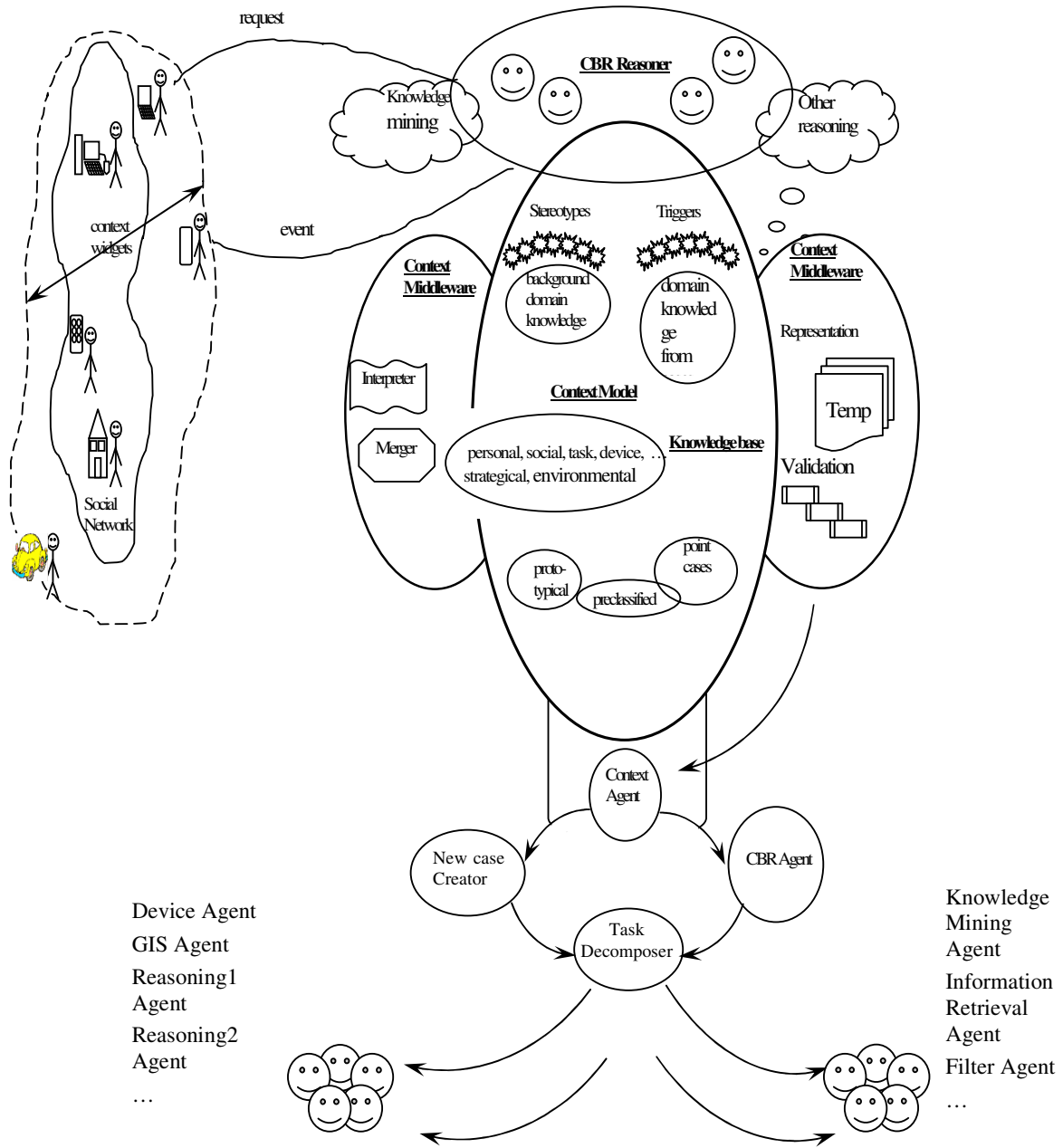

Figure 2. The ePH's multi-agent architecture

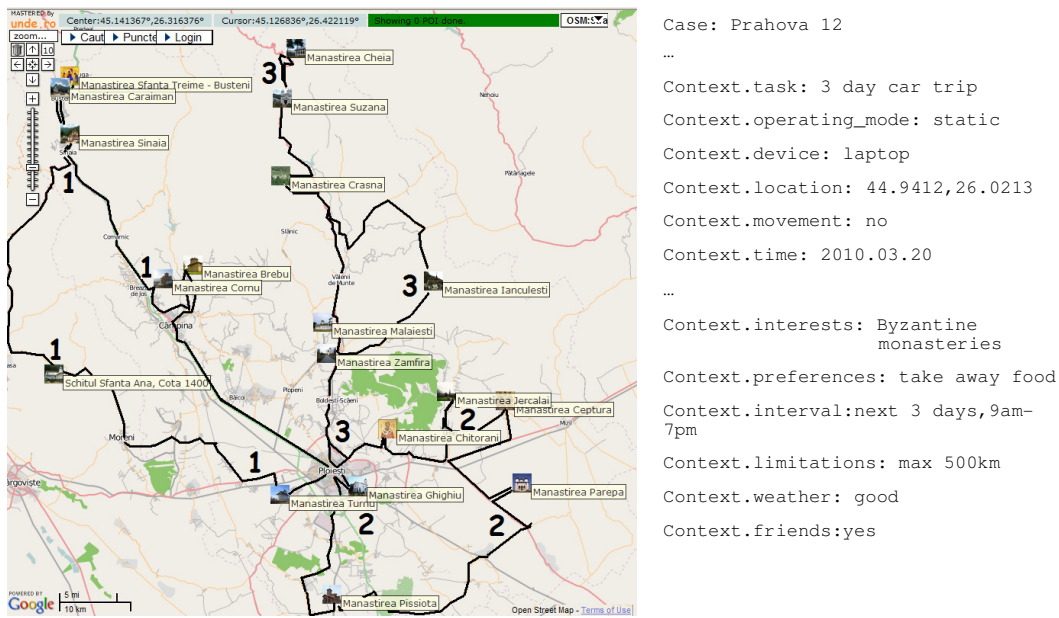

```
Case: Prahova 12
…
Context.task: 3 day car trip
Context.operating_mode: static
Context.device: laptop
Context.location: 44.9412,26.0213
Context.movement: no
Context.time: 2010.03.20
…
Context.interests: Byzantine
                   monasteries
Context.preferences: take away food
Context.interval:next 3 days,9am–
7pm
Context.limitations: max 500km
Context.weather: good
Context.friends:yes
```

Figure 3. One possible learning scenario and the related prototypical case

## 8. CONCLUSIONS AND FUTURE WORK

New technological developments have made it possible to interact with computer systems and applications anywhere and anytime. It is vital that these applications are able to adapt to the user, as a person, and to its current situation, whatever that is. Therefore, the premises for evolution towards a learning society and a knowledge economy are present. However, to have these goals accomplished requires the pervasive presence of, and ubiquitous access to, knowledge technologies that support construction of a new paradigm for learning environments and for social construction of knowledge. This new ubiquitous environment links together individuals, communities, organizations etc., both real and virtual, which are involved in impressing learning experiences based on transparent technologies and service. Learning that occur in these experiences is personalized, contextual, realistic, active, social, reflective and ubiquitous. Though, new instructional models will be necessary to cope with the challenges that this new paradigm lays before us. For example, the new learning scenarios may be too complex or the students may be overwhelmed with the support technologies, and, therefore, their learning achievements could be shallow.

In this paper we have extended the work presented in less detail in [51], and we discussed about ePH, a system that provides for sharing public interest information and knowledge that can be accessed via always-on, context-aware services, focusing on its multi-agent architecture and on its multi-dimensional context model. We have also advocated its use in learning scenarios within ubiquitous learning environments.

We need to investigate further how our agents may comply with the formal definition of an autonomous agent, which clearly distinguishes a software agent from just any program that had been proposed in [52]. Moreover, abstractization of a generic learning scenario to be used in ubiquitous learning environments is envisaged.

Future work needs to be done for better understanding of the relationship between problem solving and learning, and their integration into an autonomic framework, which provides for the system's ability to inspect its own behaviour and to learn how to change its structure, in order to improve its future performance (then the FACE will get all the parts it misses to be completely independent :), and to reach the desideratum that technology work like magic.

## REFERENCES

[1] K. Dey, (2001) "Understanding and using context", *Personal and Ubiquitous Computing*, Vol. 5, No. 1, pp. 4–7

[2] H. E. Byun, K. Cheverst, (2004) "Utilizing context history to provide dynamic adaptations", *Applied Artificial Intelligence*, Vol. 18, No. 6, pp. 533–548.

[3] C. Allison, S. A. Cerri, P. Ritrovato, A. Gaeta & M. Gaeta, (2005) "Services, Semantics, and Standards: Elements of a Learning Grid Infrastructure", *Applied Artificial Intelligence*, Vol. 19, No. 9-10, pp. 861-879

[4] M. Layte, S. Ravet, (2006) "Rethinking quality for building a learning society", *Handbook on Quality and Standardisation in E-Learning*, Springer, pp. 347-365

[5] M. Weiser, (1991) "The Computer for the Twenty-First Century", *Scientific American*, September 1991, pp. 94-100

[6] B. Bomsdorf, (2005) "Adaptation of Learning Spaces: Supporting Ubiquitous Learning in Higher Distance Education", *Dagstuhl Seminar Proceedings 05181 - Mobile Computing and Ambient Intelligence: The Challenge of Multimedia*, available online at http://drops.dagstuhl.de/opus/volltexte/2005/371, accessed October 2010

[7] G.-J. Hwang, C.-C. Tsai & S. J. H. Yang, (2008) "Criteria, Strategies and Research Issues of Context-Aware Ubiquitous Learning", *Educational Technology & Society*, Vol. 11, No. 2, pp. 81-91

[8] Y.S. Chen, T.C. Kao, J.P. Sheu & C.Y. Chiang, (2002) "A Mobile Scaffolding-Aid-Based Bird – Watching Learning System", *Proceeding of International Workshop on Wireless and Mobile Technologies in Education*, IEEE Computer Society, pp. 15-22

[9] P.S. Chiu, Y. Kuo, Y. Huang & T. Chen, (2008) "A Meaningful Learning based u-Learning Evaluation Model", *8th IEEE International Conference on Advanced Learning Technologies*, pp. 77 – 81

[10] M. Curtis, K. Luchini, W. Bobrowsky, C. Quintana & E. Soloway, (2002) "Handheld Use in K-12: A Descriptive Account", *Proceeding of the International Workshop on Wireless and Mobile Technologies in Education*, IEEE Computer Society, pp. 32-30

[11] T. de Jong, M. Specht & R. Koper, (2008). "Contextualised Media for Learning", *Educational Technology & Society*, Vol. 11, No. 2, pp. 41-53

[12] H. Ogata, Y. Yano, (2005) "Knowledge awareness for a computer-assisted language learning using handhelds", Int. J. of Continuing Engineering Education and Lifelong Learning, Vol. 14, Nos. 4/5, pp.435–449

[13] S. Yahya, E. A. Ahmad & K. A. Jalil, (2010) "The definition and characteristics of ubiquitous learning: A discussion", International *Journal of Education and Development using Information and Communication Technology (IJEDICT)*, Vol. 6, No. 1, available online at http://ijedict.dec.uwi.edu/viewarticle.php?id=785, accessed October 2010

[14] C. Yin, H. Ogata, Y. Tabata & Y. Yano, (2010) "JAPELAS2: Supporting the Acquisition of Japanese Polite Expressions in Context-Aware Ubiquitous Learning, Mobile and Ubiquitous Technologies for Language Learning", *International Journal of Mobile Learning and Organisation*, Vol. 4, No. 2, pp. 214-234

[15]    M. Vladoiu, Z. Constantinescu, (2008) "Framework for Building of a Dynamic User Community - Sharing of Context-Aware, Public Interest Information or Knowledge through Always-on Services", *10th Int'l Conf. of Enterprise Information Systems ICEIS 2008*, Barcelona, Spain, pp. 73-87

[16]    M. Vladoiu, Z. Constantinescu, (2008) "Toward Location-based Services using GPS-based Devices", *Int'l Conference on Wireless Network ICWN 2008 - World Congress on Engineering WCE 2008*, London, UK, Vol. I, pp. 799-804.

[17]    N.R. Jennings, K. Sycara & M. Wooldridge, (1998) "A roadmap of agent research and development", *Autonomous Agents and Multi-Agent Systems*, Vol. 1, No. 1, pp. 7-38.

[18]    J.-y. Hong, E.-h. Suh & S.-J. Kim, (2009) "Context-aware systems: A literature review and classification", *Expert Systems with Applications*, Vol. 36, pp. 8509–8522.

[19]    G. Chen and D. Kotz, (2000) "A Survey of Context-Aware Mobile Computing Research", *Dartmouth College*, Hanover, NH, USA, Tech. Rep. TR2000-381

[20]    A. Addis, G. Armano & E. Vargiu, (2008) "From a Generic MultiAgent Architecture to MultiAgent Information Retrieval Systems", *6th International Workshop From Agent Theory to Agent Implementation AT2AI-6*, Estoril, Portugal

[21]    W. Jih, J. Y.-j. Hsu, T.-C. Lee & L.-l. Chen, (2007) "A Multi-agent Context-aware Service Platform in a Smart Space", *Journal of Computers*, Vol. 18, No. 1, pp. 45-59

[22]    S. Ilarri, E. Mena & A. Illarramendi, (2006) "Location-dependent queries in mobile contexts: Distributed processing using mobile agents", *IEEE Trans. Mob. Comput.*, Vol. 5, No. 8, pp. 1029–1043

[23]    S. Ilarri, E. Mena & A. Illarramendi, (2010) "Location-Dependent Query Processing: Where We Are and Where We Are Heading", *ACM Computing Surveys*, Vol. 42, No. 3, pp. 1-73

[24]    G. A. Christopher, C. G. Atkeson, J. Hong, S. Long, R. Kooper & M. Pinkerton, (1997) "Cyberguide: A mobile context-aware tour guide", *Wireless Networks*, Vol. 3, No. 5, pp. 421-433

[25]    D. Camacho, R. Aler, D. Borrajo & J. M. Molina, (2005) "A Multi-Agent architecture for intelligent gathering systems", *AI Communications*, Vol. 18, No. 1, pp. 15–32

[26]    A. Kofod-Petersen, M. Mikalsen, (2005) "Context: Representation and Reasoning. Representing and Reasoning about Context in a Mobile Environment", *Revue d'Intelligence Artificielle*, Vol. 19, No. 3, pp. 479-498

[27]    Mobilearn project (2010) available online at http://www.mobilearn.org/, accessed October 2010

[28]    L. Hyon, J.-K. Yang & Y.-S. Lee, (2007) "The Development of an Ubiquitous Learning System Based On Audio Augmented Reality", *International Conference on Control, Automation and Systems*, Seoul, Korea, pp. 1072 – 1077

[29]    Y.-G. Guoa, S.-L. Wang, (2010) "Designing a Knowledge Awareness Navigation for Ubiquitous Learning Environment", available online at http://elearning.lib.fcu.edu.tw/bitstream/2377/11042/1/ce07ics002008000002.pdf, accessed October 2010

[30]    H. Ogata, C. Yin, M.M. El-Bishouty & Y. Yano, (2010) "Computer supported ubiquitous learning environment for vocabulary learning", *Int. J. of Learning Technology*, Vol. 5, No. 1, pp. 5-24

[31]    Y. Jacquinot, S. Takahashi & J. Tanaka, (2007) "Computer-assisted learning based on a ubiquitous environment - Application to Japanese Kanji learning", available online at http://www.iplab.cs.tsukuba.ac.jp/paper/domestic/yann_kccs2007.pdf, accessed October 2010

[32]    S. J. H. Yang, (2006) "Context Aware Ubiquitous Learning Environments for Peer-to-Peer Collaborative Learning", *Educational Technology & Society*, Vol. 9, No. 1, pp. 188-201

[33]    S. Joseph, K. Binsted, & D. Suthers, (2005) "PhotoStudy: Vocabulary Learning and Collaboration on Fixed and Mobile Devices", *Proceedings of IEEE International Workshop on Wireless and Mobile Technologies in Education,* Los Alamitos: IEEE Computer Society, pp. 206-210

[34]    K. Yatani, M. Sugimoto & F. Kusunoki, (2004) "Musex: a system for supporting children's collaborative learning in a museum with PDAs", *Proceedings of the 2$^{nd}$ IEEE International Workshop on Wireless and Mobile Technologies in Education*, IEEE Computer Society, pp. 109-113

[35]    I. Boticki, N. Hoic-Bozic, & I. Budiscak, (2009) "A System Architecture for a Context-aware Blended Mobile Learning Environment", Journal of Computing and Information Technology, Vol. 17, No. 2, pp. 165–175

[36]    D. A. Bergin, A. H. Anderson, T. Molnar, R. Baumgartner, S. Mitchell, S. Korper, A. Curley, & J. Rottmann, (2007) "Providing remote accessible field trips (RAFT): an evaluation study", *Computers in Human Behavior*, Vol. 23, No. 1, pp. 192-219

[37]    G.-D. Chen, & P.-Y. Chao, (2008) "Augmenting Traditional Books with Context-Aware Learning Supports from Online Learning Communities", *Educational Technology & Society*, Vol. 11, No. 2, pp. 27-40

[38]    M. Vladoiu, Z. Constantinescu, (2010) "FACE – a Knowledge-Intensive Case-Based Architecture for Context-Aware Services", *2$^{nd}$ Int'l Conference on Networked Digital Technologies NDT 2010*, Prague, Czech Republic, Springer Computer and Information Science Series - CCIS 88, pp. 533-544

[39]    Z. Constantinescu, C. Marinoiu C. & M. Vladoiu, (2010) "Driving style analysis using data mining techniques", *International Journal of Computers, Communications & Control (IJCCC)*, Vol. 5, No. 5, pp. 654-663

[40]    J. Cassens, A. Kofod-Petersen, (2006) "Using Activity Theory to Model Context Awareness: A Qualitative Case Study", *19$^{th}$ Int'l Florida AI Research Society Conference FLAIRS 2006*, Melbourne Beach, Florida, USA, pp. 619-624

[41]    P. Ozturk, A. Amodt, (1998) "A Context Model for knowledge-intensive case-based reasoning", *Int'l Journal of Human Computer Studies*, Vol. 48, pp. 331-355

[42]    E. Zibetti, V. Quera, F. S. Beltran & C. Tijus, (2001) "Contextual categorization: a mechanism linking perception and knowledge in modeling and simulating perceived events as actions", *Lecture Notes in Computer Science - LNCS 2116 - Modeling and using context*, 2001, pp. 395-408.

[43]    R. Benard, C. Bossard & P. De Loor, (2006) "Context's Modeling for Participative Simulation", in 19th Int'l Florida AI Research Society Conf. FLAIRS 2006*19$^{th}$ Int'l Florida AI Research Society Conference FLAIRS 2006*, Melbourne Beach, Florida, USA, pp. 613-618

[44]    A. Göker, H. Myrhaug, (2002) "User context and personalisation", *Workshop of the 6$^{th}$ European Conference on Case Based Reasoning ECCBR 2002*, Aberdeen, Scotland, UK, 2002, pp. 234-242

[45]    A. Aamodt, E. Plaza, (1994) "Case-Based Reasoning: Foundational Issues, Methodological Variations, and System Approaches", *AI Commun.*, Vol. 7, No. 1. , pp. 39-59

[46]    T. Chaari, E. Dejene, F. Laforest & V-M.Scuturici, (2007) "A comprehensive approach to model and use context for adapting applications in pervasive environments", *The Journal of Systems and Software*, Vol. 80, No. 12, pp. 1973-1992

[47]    A. Schmidt, C. Winterhalter, (2004) User Context Aware Delivery of E-Learning Material: Approach and Architecture, *Journal of Universal Computer Science*, Vol. 10, No. 1, pp. 38-46

[48]    L. Tankeleviciene, R. Damasevicius, (2009), Towards a Conceptual Model of Learning Context in E-learning, *Ninth IEEE International Conference on Advanced Learning Technologie, 2009 - ICALT 2009*, July, 2009, Riga, Latvia, pp. 645 – 646

[49]   A. A. Economides, (2008) "Context-aware mobile learning", The Open Knowledge Society, A Computer Science and Information Systems Manifesto, First World Summit, WSKS 2008, Athens, Greece, *Springer Communications in Computer and Information Science CCIS 19*, pp. 213-220

[50]   E. Rich, (1998) "User Modeling via Stereotypes", *Readings in intelligent user interfaces*, Morgan Kaufmann Publishers, pp. 329 – 342

[51]   M. Vladoiu, Z. Constantinescu, (2010) Learning with a Context-Aware Multiagent System, *$9^{th}$ IEEE Int'l Conference Romanian Educational Network - RoEduNet*, Sibiu, Romania, pp. 368 - 373

[52]   S. Franklin, A. Graesser, (1996) "Is it an Agent, or just a Program?: A Taxonomy for Autonomous Agents", *Workshop on Intelligent Agents III, Agent Theories, Architectures, and Languages*, Budapest, Hungary, *Lecture Notes in Computer Science - LNCS 1193*, pp. 21-35

**Authors**

Monica Vlădoiu got her MSc (1991) and PhD (2002) in the Department of Computer Science of The Polytechnic University of Bucharest, Romania. Since then, she has been with the Dept. of Informatics, Petroleum-Gas University of Ploiesti (UPG), Romania. Her main research interests include digital libraries, learning objects, multimedia databases, reflective learning, ubiquitous learning, desktop grid computing and e-society. She has published over 50 research papers concerning these topics and she has (co-) authored 3 books.

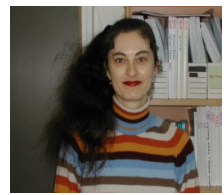

Zoran Constantinescu got his MSc (1997) in the Dept. of Computer Science of The Polytechnic University of Bucharest, Romania. Since then, he has been working both in the software engineering industry and in Higher Education. He got his doctoral degree in Computer Science (2008), from The Norwegian University of Science and Technology, Trondheim, Norway. His research interests include parallel & distributed computing, desktop grid computing, GPS systems and embedded systems. He has published over 25 research papers dealing with the above mentioned topics.

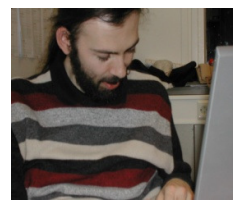